\documentclass[11pt,a4paper]{article}
\usepackage{jheppub,amsmath,amssymb,slashed,url,bm}
\usepackage{graphicx}
\usepackage{epstopdf}

\newcommand{\e}{\begin{eqnarray}}
\newcommand{\ee}{\end{eqnarray}}

\newcommand{\CN}{{\cal N}}

\def\a{\alpha}
\def\b{\beta}
\def\d{\delta}
\newcommand{\ep}{\epsilon}
\newcommand{\g}{\gamma}

\newcommand{\s}{\sigma}

\newcommand{\vp}{\varepsilon}

\font\teneurm=eurm10 \font\seveneurm=eurm7  \font\fiveeurm=eurm5
\newfam\eurmfam
\textfont\eurmfam=\teneurm \scriptfont\eurmfam=\seveneurm
\scriptscriptfont\eurmfam=\fiveeurm

\font\teneusm=eusm10 \font\seveneusm=eusm7 \font\fiveeusm=eusm5
\newfam\eusmfam
\textfont\eusmfam=\teneusm \scriptfont\eusmfam=\seveneusm
\scriptscriptfont\eusmfam=\fiveeusm

\font\tencmmib=cmmib10 \skewchar\tencmmib='177
\font\sevencmmib=cmmib7 \skewchar\sevencmmib='177
\font\fivecmmib=cmmib5 \skewchar\fivecmmib='177
\newfam\cmmibfam
\textfont\cmmibfam=\tencmmib \scriptfont\cmmibfam=\sevencmmib
\scriptscriptfont\cmmibfam=\fivecmmib

\title{A Nonabelian $(1,0)$ Tensor Multiplet Theory in 6D}

 \author{Fa-Min Chen }
\affiliation{Department of Physics, Beijing Jiaotong University, Beijing 100044, China}
\abstract{We construct a general nonabelian $(1,0)$ tensor multiplet theory in six dimensions. The gauge field of this $(1,0)$ theory is non-dynamical, and the theory contains a continuous parameter $b$.
When $b=1/2$, the $(1,0)$ theory possesses an extra discrete symmetry enhancing the supersymmetry to $(2,0)$,
and the theory
turns out to be identical to the $(2,0)$ theory of Lambert and Papageorgakis (LP). Upon dimension reduction, we obtain a general $\CN=1$ supersymmetric Yang-Mills theory in five dimensions. The applications of the theories to D4 and M5-branes are briefly discussed.}
\begin{document} \maketitle

\section{Introduction and summary}\label{secintro}
In recent years, the construction of nonabelian $(1,0)$ and $(2,0)$ superconformal theories in 6D has attracted a lot of attention. Using a Nambu 3-algebra, Lambert and Papageorgakis have been able to build up a nonabelian $(2,0)$ tensor multiplet theory in 6D \cite{LP}; In \cite{Singh1, Singh2}, the same theory has been constructed in term of ordinary Lie algebra. One particular feature of the $(2,0)$ LP theory is that the gauge field is \emph{nondynamical}. The $(2,0)$ LP theory may be a candidate of dual gauge description of interacting multiple M5-branes (a general review of superconformal field theories and multi M-branes can be found in Ref. \cite{bagger}; see also \cite{Lambert}). On the other hand, upon a dimension reduction, the $(2,0)$ LP theory can be reduced to a maximally supersymmetric Yang-Mills in 5D \cite{LP}, which can used to describe interacting multiple D4-branes.

A little later, a nonabelian $(1,0)$ theory with the same field content of a $(2,0)$ theory, has been constructed in a series of papers \cite{Sezgin1, Sezgin2, Sezgin3}; in this theory, the gauge field is \emph{dynamical}\footnote{6D theories of hypermultiplets coupled to dynamical gauge fields \emph{involving higher derivatives} were constructed in Ref. \cite{Smilga1} and studied in \cite{Simlga2, Simlga3}.}.

In this paper, we consider another possibility: we construct a nonabelian $(1,0)$ theory
with the same field content of a $(2,0)$ theory, but its gauge field is \emph{nondynamical}.
Specifically, the R-symmetry of this $(1,0)$ theory is $SU(2)$, and the theory contains
5 bosonic fields, a non-dynamical gauge field, \emph{two} anti-chiral spinor fields
(transforming in the $\textbf{2}$ of $SU(2)_L$ and $\textbf{2}$ of $SU(2)_R$ of the
global symmetry group $SU(2)_L\times SU(2)_R$, respectively), and a  selfdual field strength
 $H_{\mu\nu\rho}$. The fields are in the adjoint representation of the Lie algebra of gauge
group; and the Lie algebra of gauge symmetry can be chosen as the Lie algebra of ADE type.

One important feature of our $(1,0)$ theory is that it contains a continuous (dimensionless) parameter $b$. However, in the special case of $b=1/2$, the theory possesses an extra discrete symmetry: the theory is invariant if we exchange the two anti-chiral spinor fields; As a result, the supersymmetry can be promoted to $(2,0)$, and theory becomes identical to the $(2,0)$ LP theory. It is interesting to note that the $(1,0)$ theory with a dynamical gauge field, also contains a free parameter \cite{Sezgin3}. Perhaps this is a general feature of $(1,0)$ theory in 6D.

As in Ref. \cite{LP}, our theory also contains an auxiliary field $C^\mu$. Following the idea of \cite{LP}, if we choose the space-like vector vev $\langle C^\mu\rangle=g^2_{YM}\d^\mu_5$, the $(1,0)$ theory can be reduced to an $\CN=1$ supersymmetric Yang-Mills theory (SYM) in 5D, with $g_{YM}$ the coupling constant.
(In Section \ref{SYM}, we also briefly discuss the rest cases that $\langle C^\mu\rangle$ is a light-like or a time-like vector.)

Four dimensional $\CN=2$ SYM theories were studied intensively. For instance, in Ref. \cite{G}, a large classes of $\CN=2$ theories were constructed and studied (their gravity duals were constructed in \cite{GM}). The 5D maximally ($\CN=2$) supersymmetric Yang-Mills theories were investigated in \cite{5towers, 5kim, 5Lambert, 5Douglas}. It may be also interesting to study the 5D $\CN=1$ SYM theories constructed in this paper, and to construct their gravity duals.

The paper is organized as follows. In Section \ref{Sec10noh}, we construct a $(1,0)$ tensor multiplet theory without coupling to hypermultiplets; we then couple a $(1,0)$ hypermultiplet theory to this $(1,0)$ tensor multiplet theory in Section \ref{Sec10}. In Section \ref{20theory}, we show that the $(2,0)$ LP theory can be derived as a special case of our $(1,0)$ theory. In Section \ref{SYM}, we derive the $\CN=1$ SYM theory in 5D, starting from the $(1,0)$ theory in 6D. The applications of these theories to D4 and M5-branes are briefly discussed in Section \ref{SYM}. Our conventions and useful identities are summarized in Appendix \ref{Identities}.

\section{$(1,0)$ tensor multiplets without coupling to hypermultiplets}\label{Sec10noh}

\subsection{Closure of the superalgebra}
For simplicity, in this section we will try to construct a 6D nonabelian $(1,0)$ tensor multiplet theory with $SU(2)$ R-symmetry. This theory does not contain hypermultiplets, and its gauge field is \emph{non-dynamical}. Another motivation  is that this theory is interesting in its own right: In \cite{Sezgin1}, the $(1,0)$ tensor multiplet theory, with a \emph{dynamical} gauge field, does not contain hypermultiplets as well; It is natural to construct a similar $(1,0)$ theory with a non-dynamical gauge field, and to compare these two types of theories. Of course, in the next section, we will add the hypermultiplets into this theory such that there are non-trivial interactions between the tensor multiplets and hypermultiplets, and the field content of the final $(1,0)$ theory becomes the same as that of the $(2,0)$ theory of LP \cite{LP}.

The component fields of the $(1,0)$ tensor multiplets  in 6D are given by
\e
(\phi_m, H_{\mu\nu\rho m},\psi_{m+}),
\ee
where $\phi_m$ are the scalar fields, with $m$ an adjoint index of the Lie algebra of gauge symmetry.  The fermionic fields $\psi_{m+}$ are defined as
$\psi_{m+}=\frac{1}{2}(1+\Gamma_{6789})\psi_m$, with $\psi_m$ anti-chiral fields with respect to $\Gamma_{012345}$, i.e. $\Gamma_{012345}\psi_{m}=-\psi_{m}$ (our conventions are summarized in Appendix \ref{Identities}).  The nonabelian tensor field strengths $H_{\mu\nu\rho m}$ obey the selfdual conditions
\e\label{selfdul}
H_{\mu\nu\rho m}=\frac{1}{3!}\vp_{\mu\nu\rho\sigma\lambda\kappa}H^{\sigma\lambda\kappa}_m.
\ee
In Ref. \cite{LP}, it was proved that there are no suitable fields $B^m_{\mu\nu}$ such that $H^m_{\mu\nu\rho}=3D_{[\mu}B^m_{\nu\rho]}$ (except that the gauge group is abelian), where the covariant derivative is defined by (\ref{covd}). So in this paper, we do not try to define $H^m_{\mu\nu\rho}$ in terms of $3D_{[\mu}B^m_{\nu\rho]}$. But later we will see that $H^m_{\mu\nu\rho}$ can be solved in terms of the gauge field strength $F^m_{\mu\nu}$ (see (\ref{hf})). We will make further comments on $H^m_{\mu\nu\rho}$ below Eq. (\ref{hf}).

We postulate the law of supersymmetry transformations as follows,
\e\label{susy1}
\d\phi_m&=&-i\bar\ep_+\psi_{m+},\nonumber\\
\d\psi_{m+}&=&\Gamma^\mu\ep_+D_\mu\phi_m+\frac{1}{3!}\frac{1}{2!}
\Gamma_{\mu\nu\lambda}\ep_+H^{\mu\nu\lambda}_m,
\nonumber\\
\d A_\mu^m&=&i\bar\ep_+\Gamma_{\mu\nu}\psi^m_{+}C^{\nu}
+ic_1\bar\ep_+\psi^m_{+}C_{\mu},
\nonumber\\
\d C^\nu&=&0,\nonumber\\
\d H_{\mu\nu\rho m}&=&
3i\bar\ep_+\Gamma_{[\mu\nu}D_{\rho]}\psi_{m+}+id_1\bar\ep_{+}\Gamma_{\mu\nu\rho\sigma}
C^{\sigma}\psi_{n+}\phi_pf^{np}{}_{m},
\ee
where $C^\mu$ is an abelian auxiliary field\footnote{In a 3-algebra approach, a similar auxiliary field $C^\mu_a$ is introduced, where $a$ is a 3-algebra index \cite{LP}; Our reason for introducing $C^\mu$ into the theory is similar to that of Ref. \cite{LP} (see also \cite{Singh1, Singh2}).} with scaling dimension $-1$, and $f^{np}{}_{m}$ the structure constants of the Lie algebra. $c_1$ and $d_1$ are constants; they will be determined by the closure of superalgebra. The supersymmetry generators are defined as $\ep_{+}=\frac{1}{2}(1+\Gamma_{6789})\ep$, where $\ep$ are chiral with respect to $\Gamma_{012345}$, i.e. $\Gamma_{012345}\ep=\ep$. The covariant derivative is defined as follows
\e\label{covd}
D_\mu\phi_m=\partial_\mu\phi_m+ (A_\mu)_n\phi_pf^{np}{}_m.
\ee
Notice that in (\ref{susy1}), it is fine to replace $\psi_{m+}$ by $\psi_{m-}=\frac{1}{2}(1-\Gamma_{6789})\psi_m$, if we replace $\ep_+$ by $\ep_{-}=\frac{1}{2}(1-\Gamma_{6789})\ep$ at the same time; but the resulted theory is a \emph{new} theory.

Now we would like to examine the closure of super-Poincare algebra.
Let us begin by considering the scalar fields. A short computation gives
\e\label{scalar}
[\d_1,\d_2]\phi_m=v^\nu D_\nu\phi_m,
\ee
where
\e\label{v}
v^\mu&\equiv&-2i\bar\ep_{2+}\Gamma^\mu\ep_{1+}.
\ee
In the case of gauge fields, we obtain
\e\label{gauge}
[\d_1,\d_2]A^m_\mu&=&v^\nu F^m_{\nu\mu}-D_\mu\Lambda^m\nonumber\\
&&+v^\nu(F^m_{\mu\nu}-H^m_{\mu\nu\rho}C^\rho+2c_1C_{[\nu}D_{\mu]}\phi^m)\nonumber\\
&&-v_\mu(C^\nu D_\nu\phi^m),
\ee
where
\e
\Lambda^m&\equiv&(c_1-1)v^\nu C_\nu\phi^m,\label{lambda}\\
F^m_{\mu\nu}&=&\partial_\mu A^m_\nu-\partial_\nu A^m_\mu+[A_\mu,A_\nu]^m,
\ee
with $[A_\mu,A_\nu]^m=A_\mu^nA_\nu^pf_{np}{}^m$.
In order to close the Poincare superalgebra (up to the gauge transformation $-D_\mu\Lambda^m$), we must require the second and third lines of  (\ref{gauge}) to vanish. This gives the equations of motion of gauge fields and
the constraint equations on the scalar fields:
\e\label{EOMA}
0&=&F^m_{\mu\nu}-H^m_{\mu\nu\rho}C^\rho+2c_1C_{[\nu}D_{\mu]}\phi^m,\\
0&=&C^\nu D_\nu\phi^m.\label{constrain1}
\ee
Taking a super-variation on (\ref{constrain1}), we obtain the constraint equations on the fermionic fields
\e
0&=&C^\nu D_\nu\psi_+^m.\label{constrain2}
\ee

Note that using (\ref{lambda}), one can re-write (\ref{scalar}) as the expected form
\e\label{scalar2}
[\d_1,\d_2]\phi_m=v^\nu D_\nu\phi_m +[\Lambda, \phi]_m.
\ee

Let us now look at the auxiliary field:
\e\label{aux}
[\d_1,\d_2]C^\mu=0=v^\nu D_\nu C^\mu+[\Lambda, C^\mu].
\ee
Since we have assumed that $C^\mu$ is abelian, the commutator $[\Lambda, C^\mu]$ vanishes automatically, implying that $D_\nu C^\mu=\partial_\nu C^\mu=0$. So $C^\mu$ is actually a constant abelian field.

We now turn to the fermionic fields:
\e\label{fermion}
[\d_1,\d_2]\psi_{m+}&=&v^\nu D_\nu\psi_{m+}+[\Lambda,\psi_{+}]_m\nonumber\\
&&+\frac{3}{8}(1-c_1+d_1)v^{\mu\nu\rho ij}\Gamma_{\mu\nu\rho}\Gamma^{ij}_{-}C_\rho\psi_{n+}\phi_pf^{np}{}_{m}\nonumber\\
&&-\frac{1}{4}v^\nu D_\nu\psi_{m+}+\frac{1}{8}(7c_1-5d_1-3)v^\nu C_\nu\psi_{n+}\phi_pf^{np}{}_m
\nonumber\\
&&+\frac{1}{4}v^\nu\Gamma_{\mu\nu} D^\mu\psi_{m+}-\frac{1}{8}(c_1+d_1+3)v^\nu \Gamma_{\mu\nu}C^\mu\psi_{n+}\phi_pf^{np}{}_m,
\ee
where
\e\label{5ind}
v^{\mu\nu\rho ij}\equiv-\frac{i}{12}(\bar\ep_{2+}\Gamma^{\mu\nu\rho}
\Gamma^{ij}_-\ep_{1+}).
\ee
The second line of (\ref{fermion}) must vanish; This can be achieved by setting
\e\label{cd}
1-c_1+d_1=0.
\ee
Using (\ref{cd}), the third and fourth lines of (\ref{fermion}) become
\e
&&-\frac{1}{4}v^\nu D_\nu\psi_{m+}+\frac{1}{8}(2d_1+4)v^\nu C_\nu\psi_{n+}\phi_pf^{np}{}_m
\nonumber\\
&&+\frac{1}{4}v^\nu\Gamma_{\mu\nu} D^\mu\psi_{m+}-\frac{1}{8}(2d_1+4)v^\nu \Gamma_{\mu\nu}C^\mu\psi_{n+}\phi_pf^{np}{}_m,\nonumber\\
&=&-\frac{1}{4}v^\nu\Gamma_\nu(\Gamma^\mu D_\mu\psi_{m+}-(d_1+2)\Gamma^\mu C_\mu\psi_{n+}\phi_pf^{np}{}_m),
\ee
which leads us to impose the equations of motion
\e\label{EOMp}
0=\Gamma^\mu D_\mu\psi_{m+}-(d_1+2)\Gamma^\mu C_\mu\psi_{n+}\phi_pf^{np}{}_m.
\ee
On the other hand, taking a super-variation on (\ref{selfdul}) gives
\e\label{selfdul2}
\d H_{\mu\nu\rho m}=\frac{1}{3!}\vp_{\mu\nu\rho\sigma\lambda\kappa}\d H^{\sigma\lambda\kappa}_m.
\ee
By the last equation of (\ref{susy1}), a short computation converts the above equation into
\e\label{EOMp2}
0=\Gamma^\mu D_\mu\psi_{m+}+d_1\Gamma^\mu C_\mu\psi_{n+}\phi_pf^{np}{}_m.
\ee
Comparing the above equation and Eq. (\ref{EOMp}) determines $d_1$:
\e
d_1=-1.
\ee
Substituting $d_1=-1$ into Eq. (\ref{cd}), we learn immediately that
\e\label{c1}
c_1=0.
\ee
In other words, the second term of the third line of (\ref{susy1}) is ruled out by the closure of super-Poincare algebra.

Finally, we examine the supersymmetry transformations of the selfdual H-fields
\e\label{tensor}
&&[\d_1,\d_2]H_{\mu\nu\rho p}\nonumber\\
&=&v^\s D_\s H_{\mu\nu\rho p}+[\Lambda,H_{\mu\nu\rho}]_p\nonumber\\
&&+3(v_{[\mu}F_{\nu\rho]m}-v_{[\mu}H_{\nu\rho]\s m}C^\s)\phi_nf^{mn}{}_p\\
&&+4v^\s[D_{[\mu}H_{\nu\rho\s]p}+\frac{i}{8}\vp_{\mu\nu\rho\lambda\s\tau}(\bar\psi_{m+}\Gamma^\tau
\psi_{n+})C^\lambda f^{mn}{}_p+\frac{1}{4}\vp_{\mu\nu\rho\lambda\s\tau}\phi_mD^\tau\phi_nC^\lambda f^{mn}{}_p].\nonumber
\ee
The second line vanishes on account of Eqs. (\ref{EOMA}) and (\ref{c1}), while the second line must be the equations of motion
\e\label{EOMH}
0=D_{[\mu}H_{\nu\rho\s]p}+\frac{i}{8}\vp_{\mu\nu\rho\lambda\s\tau}(\bar\psi_{m+}\Gamma^\tau
\psi_{n+})C^\lambda f^{mn}{}_p+\frac{1}{4}\vp_{\mu\nu\rho\lambda\s\tau}\phi_mD^\tau\phi_nC^\lambda f^{mn}{}_p.
\ee
Using Eqs. (\ref{EOMA}) and (\ref{EOMH}), and the Bianchi identity $D_{[\mu}F_{\nu\rho]p}=0$, one obtains the constraint equations
\e\label{eqH}
C^\s D_\s H_{\mu\nu\rho p}=0.
\ee

The equations of motion for the scalar fields can be obtained by taking a super-variation on (\ref{EOMp2}); they are given by
\e\label{EOMS}
0&=&D^2\phi_p-\frac{i}{2}(\bar\psi_{m+}\Gamma_\nu\psi_{n+})C^\nu f^{mn}{}_p.
\ee

\subsection{Summary of the theory}\label{Secsumm1}

In summary, the equations of the $(1,0)$ theory of this section are given by
\e\label{eqs1}
0&=&D^2\phi_p-\frac{i}{2}(\bar\psi_{m+}\Gamma_\nu\psi_{n+})C^\nu f^{mn}{}_p,
\nonumber\\
0&=&F^m_{\mu\nu}-H^m_{\mu\nu\rho}C^\rho\nonumber\\
0&=&\Gamma^\mu D_\mu\psi_{m+}-\Gamma^\mu C_\mu\psi_{n+}\phi_pf^{np}{}_m
\nonumber\\
0&=&D_{[\mu}H_{\nu\rho\s]p}+\frac{i}{8}\vp_{\mu\nu\rho\lambda\s\tau}(\bar\psi_{m+}\Gamma^\tau
\psi_{n+})C^\lambda f^{mn}{}_p+\frac{1}{4}\vp_{\mu\nu\rho\lambda\s\tau}\phi_mD^\tau\phi_nC^\lambda f^{mn}{}_p.
\nonumber\\
0&=&C^\s D_\s\phi_{m}=C^\s D_\s\psi_{m+}=C^\s D_\s H_{\mu\nu\rho m}=\partial_\mu C^\nu.
\ee
Multiplying both sides of the second equation by $C^\nu$ gives
\e\label{massless}
F^m_{\mu\nu}C^\nu=0.
\ee
Notice that the second equation is
equivalent to the equation
\e\label{hf}
C^2H^m_{\mu\nu\rho}=3F^m_{[\mu\nu}C_{\rho]}+\frac{1}{2}\vp_{\mu\nu\rho}{}^{\lambda\kappa\tau}F^m_{\lambda\kappa}C_\tau.
\ee

We emphasize again that the gauge field of the $(1,0)$ theory of this section is nondynamical; It seems that due to this reason, one cannot define the nonabelian selfdual field strength in the following way:
$H_{\mu\nu\rho m}=3D_{[\mu}B_{\nu\rho]m}$, as analyzed by the authors of \cite{LP}. However, in the $(1,0)$ theory of \cite{Sezgin1}, the gauge field is dynamical, and it is possible to construct a field strength $H_{\mu\nu\rho m}$ associated with the nonabelian covariant derivative $D_{[\mu}B_{\nu\rho]m}$ (see also \cite{HO, Chu, fss, chu2, HO2, Huang, chu3, chu4, Saemann1}). So these two types of theories are not the same; But still, there may be a connection between them. It would be interesting to investigate these two types of theories further.

The supersymmetry transformations are
\e\label{susy2}
\d\phi_m&=&-i\bar\ep_+\psi_{m+},\nonumber\\
\d\psi_{m+}&=&\Gamma^\mu\ep_+D_\mu\phi_m+\frac{1}{3!}\frac{1}{2!}
\Gamma_{\mu\nu\lambda}\ep_+H^{\mu\nu\lambda}_m,
\nonumber\\
\d A_\mu^m&=&i\bar\ep_+\Gamma_{\mu\nu}\psi^m_{+}C^{\nu},
\nonumber\\
\d C^\nu&=&0,\nonumber\\
\d H_{\mu\nu\rho m}&=&
3i\bar\ep_+\Gamma_{[\mu\nu}D_{\rho]}\psi_{m+}-i\bar\ep_{+}\Gamma_{\mu\nu\rho\sigma}
C^{\sigma}\psi_{n+}\phi_pf^{np}{}_{m}.
\ee
If we make the following replacements
\e
\ep_+\rightarrow \ep_-,\quad \psi_{m+}\rightarrow\psi_{m-}
\ee
in (\ref{eqs1}) and (\ref{susy2}), we will obtain a \emph{new} $(1,0)$ theory.

\section{$(1,0)$ tensor multiplets coupling to hypermultiplets}\label{Sec10}

\subsection{$(1,0)$ tensor multiplets coupling to hypermultiplets}
Having constructed the $(1,0)$ tensor multiplet theory, our next challenge is to couple this theory to a $(1,0)$ hypermultiplet theory. We begin by considering the free $(1,0)$ hypermultiplet  $(X^i, \psi_{-})$. Here $X^i$ ($i=6,7,8,9$) are a set of bosonic fields; And the ferminoic field $\psi_-$ is anti-chiral with respect to  $\Gamma_{6789}$ as well as $\Gamma_{012345}$, that is, $\psi_-=\frac{1}{2}(1-\Gamma_{6789})\psi$ and $\Gamma_{012345}\psi=-\psi$ (our conventions are summarized in Section \ref{Identities}). The $(1,0)$ supersymmetry transformations are
\e
\d X^i&=&i\ep_+\Gamma^i\psi_-,\nonumber\\
\d\psi_-&=&\Gamma^\mu\Gamma^i\ep_+\partial_\mu X^i.
\ee
The superalgebra is closed by imposing the equations of motion
\e\label{freepsi}
\Gamma^\mu\partial_\mu\psi_-=0.
\ee
Taking a super-variation on (\ref{freepsi}) gives the equations of motion of the free bosonic fields:
$\partial^\mu\partial_\mu X^i=0$. Clearly, this free $(1,0)$ hypermultiplet theory has an $SU(2)$ R-symmetry.

To couple tensor multiplets and hypermultiplets, it is natural to assume that they
share the \emph{same} gauge symmetry, and the hypermultilets furnish the adjoint representation of the algebra
of gauge symmetry, as the tensor multiplets do. Under these assumptions, 
we propose the supersymmetry transformations
\e\label{susy3}
\d\phi_m&=&-i\bar\ep_+\psi_{m+},\nonumber\\
\d X^i_m&=&i\bar\ep_+\Gamma^i\psi_{m-},\nonumber\\
\d\psi_{m-}&=&\Gamma^\mu\Gamma^i\ep_+D_\mu X^i_m+a\Gamma_\lambda\Gamma^i\ep_+ C^{\lambda } X^i_n\phi_p f^{np}{}_{m},\nonumber\\
\d\psi_{m+}&=&\Gamma^\mu\ep_+D_\mu\phi_m+\frac{1}{3!}\frac{1}{2!}
\Gamma_{\mu\nu\lambda}\ep_+H^{\mu\nu\lambda}_m+b\Gamma_\lambda\Gamma^{ij}_{-}\ep_+ C^{\lambda } X^i_nX^j_p f^{np}{}_{m},
\nonumber\\
\d A_\mu^m&=&i\bar\ep_+\Gamma_{\mu\nu}\psi^m_{+}C^{\nu},
\nonumber\\
\d C^\nu&=&0,\nonumber\\
\d H_{\mu\nu\rho m}&=&
3i\bar\ep_+\Gamma_{[\mu\nu}D_{\rho]}\psi_{m+}-i\bar\ep_{+}\Gamma_{\mu\nu\rho\sigma}
C^{\sigma}\psi_{n+}\phi_pf^{np}{}_{m}\nonumber\\
&&+id\bar\ep_{+}\Gamma^i\Gamma_{\mu\nu\rho\sigma}\psi_{n-}C^{\sigma
}X^i_pf^{np}{}_{m},
\ee
where $a$, $b$, and $d$ are constants, to be determined later. We see that after coupling to the hypermultiplets, the field content of this $(1,0)$ theory is the same as that of the $(2,0)$ theory of LP \cite{LP}. However, the R-symmetry of the $(1,0)$ theory is only $SU(2)$, while the R-symmetry of the $(2,0)$ theory is $SO(5)$.

Let us now examine the closure of the superalgebra. Using the results of the last section, the task of examining the closure of the algebra becomes much simpler. The transformation of scalar fields $\phi_m$ remains the same form as that of the last section:
\e\label{scalar2}
[\d_1,\d_2]\phi_m=v^\nu D_\nu\phi_m+[\Lambda, \phi]_m.
\ee
The set of parameters of gauge transformation $\Lambda^m=-v^\nu C_\nu\phi^m$ are also unchanged (see (\ref{lambda}) and (\ref{c1})), with $v^\nu$ defined by Eq. (\ref{v}). Also, the commutator $[\d_1,\d_2]C^\mu$ remains the same as Eq. (\ref{aux}).

In the case of bosonic fields $X^i_m$, a short calculation gives
\e\label{bosonic2}
[\d_1,\d_2]X^i_m=v^\nu D_\nu X^i_m+a[\Lambda, X^i]_m.
\ee
It can be seen that we must set
\e
a=1.
\ee

Using the identity (\ref{Fierz2}), we see that the commutator $[\d_1,\d_2]A^m_\mu$ also remains the same form as that of (\ref{gauge}):
\e\label{gauge2}
[\d_1,\d_2]A^m_\mu&=&v^\nu F^m_{\nu\mu}-D_\mu\Lambda^m\nonumber\\
&&+v^\nu(F^m_{\mu\nu}-H^m_{\mu\nu\rho}C^\rho)\nonumber\\
&&-v_\mu(C^\nu D_\nu\phi^m).
\ee
The last two lines must vanish separately. In this way, we obtain the equations of motion of gauge fields and the constraint equations on the scalar fields $\phi_m$:
\e
0&=&F^m_{\mu\nu}-H^m_{\mu\nu\rho}C^\rho,\\
0&=&C^\nu D_\nu\phi^m.
\ee

Using the identities in Appendix \ref{Identities}, one obtains
\e\label{fermion2m}
[\d_1,\d_2]\psi_{m-}&=&v^\nu D_\nu\psi_{m-}+[\Lambda,\psi_{-}]_m\nonumber\\
&&-\frac{1}{2}v^\nu\Gamma_\nu(\Gamma^\mu D_\mu\psi_{m-}+\Gamma^\mu C_\mu\psi_{n-}\phi_pf^{np}{}_m+\Gamma^\mu \Gamma^i C_\mu\psi_{n+}X^i_pf^{np}{}_m).
\ee
We see that the second line must be equations of motion
\e\label{EOMp3}
0&=&\Gamma^\mu D_\mu\psi_{m-}+\Gamma^\mu C_\mu\psi_{n-}\phi_pf^{np}{}_m+\Gamma^\mu \Gamma^i C_\mu\psi_{n+}X^i_pf^{np}{}_m.
\ee

As for $\psi_{m+}$, we have
\e\label{fermion2p}
[\d_1,\d_2]\psi_{m+}&=&v^\nu D_\nu\psi_{m+}+[\Lambda,\psi_{+}]_m\nonumber\\
&&-\frac{1}{4}v^\nu\Gamma_\nu(\Gamma^\mu D_\mu\psi_{m+}-\Gamma^\mu C_\mu\psi_{n+}\phi_pf^{np}{}_m+d\Gamma^\mu\Gamma^iC_\mu\psi_{n-}X^i_pf^{np}{}_m)\nonumber\\
&&+\frac{3}{8}(2b-d)v^\nu C^\mu\Gamma_\mu\Gamma_\nu\Gamma^i\psi_{n-}X^i_pf^{np}{}_m
\nonumber\\
&&-\frac{3}{2}(2b-d)v^{\mu\nu\rho ij}\Gamma_{\mu\nu}\Gamma^iC_\rho\psi_{n-}X^j_pf^{np}{}_m,
\ee
where $v^{\mu\nu\rho ij}$ is defined by (\ref{5ind}). The first line and the first two terms of second line are the results of Section \ref{Sec10noh} (see (\ref{fermion})). The rest terms are due to that we have coupled the $(1,0)$ theory of Section \ref{Sec10noh} to the $(1,0)$ hypermultiplet theory. It can be seen immediately that if
\e\label{d}
d=2b,
\ee
the last two lines vanish simultaneously. Thus the second line of (\ref{fermion2p}) must be the equations of motion
\e\label{EOMp4}
0&=&\Gamma^\mu D_\mu\psi_{m+}-\Gamma^\mu C_\mu\psi_{n+}\phi_pf^{np}{}_m+2b\Gamma^\mu\Gamma^iC_\mu\psi_{n-}X^i_pf^{np}{}_m.
\ee
The above equation can be also derived by combining the selfdual conditions
$\d H_{\mu\nu\rho m}=\frac{1}{3!}\vp_{\mu\nu\rho\sigma\lambda\kappa}\d H^{\sigma\lambda\kappa}_m$ and Eq. (\ref{d}).

Let us now compute the super-variation of the H-fields. After some algebraic steps, we obtain
\e\label{tensor2}
&&[\d_1,\d_2]H_{\mu\nu\rho p}\nonumber\\
&=&v^\s D_\s H_{\mu\nu\rho p}+[\Lambda,H_{\mu\nu\rho}]_p\nonumber\\
&&+3(v_{[\mu}F_{\nu\rho]m}-v_{[\mu}H_{\nu\rho]\s m}C^\s)\phi_nf^{mn}{}_p\\
&&+4v^\s[D_{[\mu}H_{\nu\rho\s]p}+\frac{i}{8}\vp_{\mu\nu\rho\lambda\s\tau}(\bar\psi_{m+}\Gamma^\tau
\psi_{n+})C^\lambda f^{mn}{}_p+\frac{1}{4}\vp_{\mu\nu\rho\lambda\s\tau}\phi_mD^\tau\phi_nC^\lambda f^{mn}{}_p\nonumber\\
&&\quad\quad\quad+\frac{i}{4}b\vp_{\mu\nu\rho\lambda\s\tau}(\bar\psi_{m-}\Gamma^\tau
\psi_{n-})C^\lambda f^{mn}{}_p+\frac{1}{2}b\vp_{\mu\nu\rho\lambda\s\tau}X^i_mD^\tau X^i_nC^\lambda f^{mn}{}_p]\nonumber\\
&&-48bv_{\mu\nu\rho}{}^{ij}(C^\s D_\s X^i_m)X^j_nf^{mn}{}_p.\nonumber
\ee
The first three lines are adopted from (\ref{tensor}), while the last two lines are due to the coupling of $(1,0)$ tensor multiplet and hypermultiplet theories. To close the algebra, we must require the last line to vanish. This can be done by either setting $b=0$ or $C^\s D_\s X^i_m=0$. However, if $b=0$, there wouldn't be nontrivial interactions between the tensor multiplets and hypermultiplets. We are therefore led to
\e\label{eqboso}
C^\s D_\s X^i_m=0.
\ee
The second line of (\ref{tensor2}) goes away by the equations of motion for the gauge fields, while the third and fourth lines must be the equations of motion of the H-fields
\e
0&=&D_{[\mu}H_{\nu\rho\s]p}+\frac{i}{8}\vp_{\mu\nu\rho\lambda\s\tau}(\bar\psi_{m+}\Gamma^\tau
\psi_{n+})C^\lambda f^{mn}{}_p+\frac{1}{4}\vp_{\mu\nu\rho\lambda\s\tau}\phi_mD^\tau\phi_nC^\lambda f^{mn}{}_p\nonumber\\
&&+\frac{i}{4}b\vp_{\mu\nu\rho\lambda\s\tau}(\bar\psi_{m-}\Gamma^\tau
\psi_{n-})C^\lambda f^{mn}{}_p+\frac{1}{2}b\vp_{\mu\nu\rho\lambda\s\tau}X^i_mD^\tau X^i_nC^\lambda f^{mn}{}_p.
\ee

Taking a super-variation on (\ref{eqboso}) gives the constraint equations
\e\label{eqpsim}
C^\s D_\s \psi_{m-}=0.
\ee
Also, in exactly the same way of deriving Eq. (\ref{eqH}), one can obtain the constraint equations
\e\label{eqH2}
C^\s D_\s H_{\mu\nu\rho p}=0.
\ee

Finally, one can derive the equations of motion of the bosonic fields $X^i_p$ and $\phi_p$ by taking super-variations on Eqs. (\ref{EOMp3}) and (\ref{EOMp4}), respectively; they are given by
\e
0&=&D^2X^i_p+i(\bar\psi_{m+}\Gamma_\nu\Gamma^i\psi_{n-})C^\nu f^{mn}{}_p
-C^2(\phi_mX^i_n\phi_q+2bX^j_mX^i_nX^j_q)f^{mn}{}_of^{oq}{}_p,
\\
0&=&D^2\phi_p-\frac{i}{2}[(\bar\psi_{m+}\Gamma_\nu\psi_{n+})
-2b(\bar\psi_{m-}\Gamma_\nu\psi_{n-})]C^\nu f^{mn}{}_p-2bC^2X^i_m\phi_nX^i_qf^{mn}{}_of^{oq}{}_p.\nonumber
\ee
In deriving the above equations, we have used the identities (\ref{Fierz3}) and (\ref{Fierz4}).

We see that the continuous parameter $b$ survives in the $(1,0)$ theory: it cannot be fixed by the closure of superalgebra, and it cannot be absorbed into the definitions of fields, either. In Section \ref{20theory} we will see that in the special case of $b=\frac{1}{2}$, the $(1,0)$ supersymmetry can be enhanced to $(2,0)$.
\subsection{Summary of the theory}\label{sec10th}
In summary, the equations of the $(1,0)$ theory of this section are
\e\label{equations2}
0&=&D^2X^i_p+i(\bar\psi_{m+}\Gamma_\nu\Gamma^i\psi_{n-})C^\nu f^{mn}{}_p
-C^2(\phi_mX^i_n\phi_q+2bX^j_mX^i_nX^j_q)f^{mn}{}_of^{oq}{}_p,
\nonumber\\
0&=&D^2\phi_p-\frac{i}{2}[(\bar\psi_{m+}\Gamma_\nu\psi_{n+})
-2b(\bar\psi_{m-}\Gamma_\nu\psi_{n-})]C^\nu f^{mn}{}_p-2bC^2X^i_m\phi_nX^i_qf^{mn}{}_of^{oq}{}_p,
\nonumber\\
0&=&F^m_{\mu\nu}-H^m_{\mu\nu\rho}C^\rho,\nonumber\\
0&=&\Gamma^\mu D_\mu\psi_{m-}+\Gamma^\mu C_\mu\psi_{n-}\phi_pf^{np}{}_m+\Gamma^\mu \Gamma^i C_\mu\psi_{n+}X^i_pf^{np}{}_m,\nonumber\\
0&=&\Gamma^\mu D_\mu\psi_{m+}-\Gamma^\mu C_\mu\psi_{n+}\phi_pf^{np}{}_m+2b\Gamma^\mu\Gamma^iC_\mu\psi_{n-}X^i_pf^{np}{}_m,\\
0&=&D_{[\mu}H_{\nu\rho\s]p}+\frac{i}{8}\vp_{\mu\nu\rho\lambda\s\tau}[(\bar\psi_{m+}\Gamma^\tau
\psi_{n+})+2b(\bar\psi_{m-}\Gamma^\tau
\psi_{n-})]C^\lambda f^{mn}{}_p\nonumber\\
&&+\frac{1}{4}\vp_{\mu\nu\rho\lambda\s\tau}(\phi_mD^\tau\phi_n+2bX^i_mD^\tau X^i_n)C^\lambda f^{mn}{}_p,
\nonumber\\
0&=&C^\s D_\s\phi_{m}=C^\s D_\s X^i_{m}
=C^\s D_\s\psi_{m-}=C^\s D_\s\psi_{m+}=C^\s D_\s H_{\mu\nu\rho m}=\partial_\mu C^\nu.\nonumber
\ee

The supersymmetry transformations are
\e\label{susy4}
\d\phi_m&=&-i\bar\ep_+\psi_{m+},\nonumber\\
\d X^i_m&=&i\bar\ep_+\Gamma^i\psi_{m-},\nonumber\\
\d\psi_{m-}&=&\Gamma^\mu\Gamma^i\ep_+D_\mu X^i_m+\Gamma_\lambda\Gamma^i\ep_+ C^{\lambda } X^i_n\phi_p f^{np}{}_{m},\nonumber\\
\d\psi_{m+}&=&\Gamma^\mu\ep_+D_\mu\phi_m+\frac{1}{3!}\frac{1}{2!}
\Gamma_{\mu\nu\lambda}\ep_+H^{\mu\nu\lambda}_m+b\Gamma_\lambda\Gamma^{ij}_{-}\ep_+ C^{\lambda } X^i_nX^j_p f^{np}{}_{m},
\nonumber\\
\d A_\mu^m&=&i\bar\ep_+\Gamma_{\mu\nu}\psi^m_{+}C^{\nu},
\nonumber\\
\d C^\nu&=&0,\nonumber\\
\d H_{\mu\nu\rho m}&=&
3i\bar\ep_+\Gamma_{[\mu\nu}D_{\rho]}\psi_{m+}-i\bar\ep_{+}\Gamma_{\mu\nu\rho\sigma}
C^{\sigma}\psi_{n+}\phi_pf^{np}{}_{m}\nonumber\\
&&+2ib\bar\ep_{+}\Gamma^i\Gamma_{\mu\nu\rho\sigma}\psi_{p-}C^{\sigma
}X^i_nf^{np}{}_{m}.
\ee

In constructing the $(1,0)$ theory, we have used $\ep_+$
as the set of generators of supersymmetry. The generators $\ep_+$ transform as $\textbf{2}$ of $SU(2)_L$ of the global symmetry group $SO(4)=SU(2)_L\times SU(2)_R$. We can of course choose $\ep_-$, transforming as $\textbf{2}$ of $SU(2)_R$, as the set of supersymmetry generators. In fact, if we make the replacement
\e
\ep_+\rightarrow \ep_-
\ee
in (\ref{susy4}), and switch $\psi_{m+}$ and $\psi_{m-}$ in (\ref{susy4}) and (\ref{equations2})
\e\label{replace}
\psi_{m+}\leftrightarrow \psi_{m-},
\ee
we will obtain a \emph{new} $(1,0)$ theory, provided that $b\neq\frac{1}{2}$, because the discrete transformation (\ref{replace}) is \emph{not} a symmetry of the theory (see (\ref{EOMp5}) and the comments below (\ref{EOMp5})).


The gauge groups can be classified by specifying the structure constants $f^{mn}{}_{p}$ and the invariant symmetric tensor $k_{mn}$ on the Lie algebras. (In the case of simple or semi-simple Lie algebra, $k_{mn}$ is the Killing-Cartan metric.) For instance, the Lie algebras can be chosen
as the Lie algebras of type ADE.

Finally, we expect that the $(1,0)$ theory has a full $OSp(8|2)$ suerconformal symmetry. One should be able to verify this symmetry explicitly. The ideas for proving the $OSp(N|4)$ ($N=4, 5, 6, 8$) superconformal symmetries associated with the 3D $\CN\geq4$ Chern-Simons matter theories may be useful \cite{Schwarz, Schwarz0, chen8, chen9}.

\subsection{Promoting to $(2,0)$ LP theory}\label{20theory}
Since a theory with fewer supersymmetries must be more general than those with higher supersymmetries, we should be able to obtain the $(2,0)$ theory as a special case of the $(1,0)$ theory. To enhance the supersymmetry is equivalent to promote the $SU(2)$ R-symmetry to $SO(5)$. The first step is to require the bosonic fields to transform as the $\textbf{5}$ of $SO(5)$; this leads us to define
\e
X^I=(X^i, \phi),
\ee
where $i=6,7,8,9$ and $I=6,\ldots,10$. Similarly, we can combine $\Gamma^i$ and $\Gamma^{10}=\Gamma_{012345}\Gamma_{6789}$ to form
\e
\Gamma^I=(\Gamma^i, \Gamma^{10}).
\ee
With these notations and the properties of $\ep_+$, $\psi_{m+}$, and $\psi_{m-}$ (see (\ref{eppsi})), the fourth and fifth equations of (\ref{equations2}) can be unified into the equation
\e\label{EOMp5}
0&=&\Gamma^\mu D_\mu\psi_m+\Gamma^\mu\Gamma^IC_\mu\psi_nX^I_pf^{np}{}_{m}
+(2b-1)\Gamma^\mu\Gamma^iC_\mu\psi_{n-}X^i_pf^{np}{}_m,
\ee
where $\psi_{m}=\psi_{m+}+\psi_{m-}$. It can be seen that under the discrete transformation $\psi_{m+}\leftrightarrow\psi_{m-}$, the first two terms of (\ref{EOMp5}) are invariant, but the last term becomes
\e
(2b-1)\Gamma^\mu\Gamma^iC_\mu\psi_{n+}X^i_pf^{np}{}_m.
\ee
Hence the transformation is not a symmetry of the theory in general. However, in the special case of
\e\label{b}
b=\frac{1}{2},
\ee
the last term of (\ref{EOMp5}) drops, and equation (\ref{EOMp5}) is invariant under the transformation $\psi_{m+}\leftrightarrow\psi_{m-}$; more importantly, this equation becomes manifestly $SO(5)$ covariant. Similarly, using $b=1/2$, the first two equations of (\ref{equations2}) can be combined into a single equation with manifest $SO(5)$ symmetry. The rest equations of (\ref{equations2}) can be taken care of as well.

In summary, if $b=1/2$, we have
\e\label{equations3}
0&=&D^2X^I_p+\frac{i}{2}(\bar\psi_{m}\Gamma_\nu\Gamma^I\psi_{n})C^\nu f^{mn}{}_p
-C^2X^J_mX^I_nX^J_qf^{mn}{}_of^{oq}{}_p,
\nonumber\\
0&=&F^m_{\mu\nu}-H^m_{\mu\nu\rho}C^\rho,\nonumber\\
0&=&\Gamma^\mu D_\mu\psi_m+\Gamma^\mu\Gamma^IC_\mu\psi_nX^I_pf^{np}{}_{m},\\
0&=&D_{[\mu}H_{\nu\rho\s]p}+\frac{i}{8}\vp_{\mu\nu\rho\lambda\s\tau}(\bar\psi_{m}\Gamma^\tau
\psi_{n})C^\lambda f^{mn}{}_p+\frac{1}{4}\vp_{\mu\nu\rho\lambda\s\tau}X^I_mD^\tau X^I_nC^\lambda f^{mn}{}_p,
\nonumber\\
0&=&C^\s D_\s X^I_{m}
=C^\s D_\s\psi_{m}=C^\s D_\s H_{\mu\nu\rho m}=\partial_\mu C^\nu.\nonumber
\ee
These are essentially the same equations of the $(2,0)$ LP theory \footnote{The set of equations for this $(2,0)$ theory can be also developed via twistor techniques \cite{Saemann2, Saemann3}.}, constructed in terms of 3-algebra \cite{LP}. Similarly, if $b=1/2$, the supersymmetry transformations (\ref{susy4}) can be recast into the forms
\e\label{susy5}
\d X^I_m&=&i\bar\ep_{+}\Gamma^I\psi_{m},\nonumber\\
\d\psi_{m}&=&\Gamma^\mu\Gamma^I\ep_{+} D_\mu X^I_m+\frac{1}{3!}\frac{1}{2!}
\Gamma_{\mu\nu\lambda}\ep_{+} H^{\mu\nu\lambda}_m+\frac{1}{2}\Gamma_\lambda\Gamma^{IJ}
\ep_{+} C^{\lambda } X^I_nX^J_p f^{np}{}_{m},
\nonumber\\
\d A_\mu^m&=&i\bar\ep_{+}\Gamma_{\mu\nu}\psi^mC^{\nu},
\nonumber\\
\d C^\nu&=&0,\nonumber\\
\d H_{\mu\nu\rho m}&=&
3i\bar\ep_{+}\Gamma_{[\mu\nu}D_{\rho]}\psi_{m}
+i\bar\ep_{+}\Gamma^I\Gamma_{\mu\nu\rho\sigma}\psi_{n}C^{\sigma
}X^I_pf^{np}{}_{m}.
\ee

At the end of Section (\ref{sec10th}), we asserted that we can obtain a new $(1,0)$ theory by  making the replacement $\ep_+\rightarrow\ep_-$ and the transformation $\psi_{m+}\leftrightarrow \psi_{m-}$. However, this is not the case if $b=\frac{1}{2}$. Applying
\e
\ep_+\rightarrow\ep_-,\quad\psi_{m+}\leftrightarrow \psi_{m-}\label{trans}
\ee
to Eqs. (\ref{equations3}) and (\ref{susy5}), we find that (1) the equations of motion (\ref{equations3}) do not change at all, meaning that $\psi_{m+}\leftrightarrow \psi_{m-}$ is just a discrete symmetry of the theory; (2) the supersymmetry transformations (\ref{susy5})
become
\e\label{susy5m}
\d X^I_m&=&i\bar\ep_{-}\Gamma^I\psi_{m},\nonumber\\
\d\psi_{m}&=&\Gamma^\mu\Gamma^I\ep_{-} D_\mu X^I_m+\frac{1}{3!}\frac{1}{2!}
\Gamma_{\mu\nu\lambda}\ep_{-} H^{\mu\nu\lambda}_m+\frac{1}{2}\Gamma_\lambda\Gamma^{IJ}
\ep_{-} C^{\lambda } X^I_nX^J_p f^{np}{}_{m},
\nonumber\\
\d A_\mu^m&=&i\bar\ep_{-}\Gamma_{\mu\nu}\psi^mC^{\nu},
\nonumber\\
\d C^\nu&=&0,\nonumber\\
\d H_{\mu\nu\rho m}&=&
3i\bar\ep_{-}\Gamma_{[\mu\nu}D_{\rho]}\psi_{m}
+i\bar\ep_{-}\Gamma^I\Gamma_{\mu\nu\rho\sigma}\psi_{n}C^{\sigma
}X^I_pf^{np}{}_{m},
\ee
which must be considered as another independent set of supersymmetry transformations  of the theory.  The two sets of supersymmetry transformations (\ref{susy5}) and (\ref{susy5m}) can be unified into
\e\label{susy6}
\d X^I_m&=&i\bar\ep\Gamma^I\psi_{m},\nonumber\\
\d\psi_{m}&=&\Gamma^\mu\Gamma^I\ep D_\mu X^I_m+\frac{1}{3!}\frac{1}{2!}
\Gamma_{\mu\nu\lambda}\ep H^{\mu\nu\lambda}_m+\frac{1}{2}\Gamma_\lambda\Gamma^{IJ}
\ep C^{\lambda } X^I_nX^J_p f^{np}{}_{m},
\nonumber\\
\d A_\mu^m&=&i\bar\ep\Gamma_{\mu\nu}\psi^mC^{\nu},
\nonumber\\
\d C^\nu&=&0,\nonumber\\
\d H_{\mu\nu\rho m}&=&
3i\bar\ep\Gamma_{[\mu\nu}D_{\rho]}\psi_{m}
+i\bar\ep\Gamma^I\Gamma_{\mu\nu\rho\sigma}\psi_{n}C^{\sigma
}X^I_pf^{np}{}_{m},
\ee
where $\ep=\ep_++\ep_-$. These are essentially the same supersymmetry transformations of the $(2,0)$ LP theory, expressed in terms of Lorentian 3-algebra \cite{LP}. In this way, we have recovered
the whole $(2,0)$ LP theory.

\subsection{Relating to $\CN=1$ SYM in 5D}\label{SYM}
In this section, we will show our (1,0) theory can be reduced to an $\CN=1$ super Yang-Mills theory in 5D by specifying the auxiliary field $C^\mu$. Since our $C^\mu$
plays the same role as that of $C^\mu_a$ of the LP theory \cite{LP} (see also \cite{Singh1, Singh2}), it is natural adopt the method in \cite{LP} and to choose the space-like vector vev
\e\label{vev}
\langle C^\mu\rangle=g(0,\ldots,0,1)=g\d^\mu_5,
\ee
where the constant $g$ has dimension $-1$ and obeys the equation $\partial_\nu g=0$. The equations of motion of gauge fields (the 3rd equation of (\ref{equations2})) now become
\e
F_{\a\b m}=gH_{\a\b5m},
\ee
where we have decomposed $\mu$ into $\mu=({\a, 5})$, with $\a=0, 1,\ldots,4.$ On the other hand, since $F_{5\b m}=gH_{5\b5m}=0$, we have
\e
F_{5\b}=\partial_5A_\b-\partial_\b A_5+[A_5,A_\b]=0.
\ee
So locally, we may set the flat connection $A_5=0$. As a result, $A_\b$ is independent of the fifth coordinate $x^5$, i.e. $\partial_5A_\b=0$.

The rest equations in (\ref{equations2}) can be reduced to those of $\CN=1$ SYM theory in 5D
\e\label{equations4}
0&=&D^\a D_\a X^i_p+ig(\bar\psi_{m+}\Gamma_5\Gamma^i\psi_{n-})f^{mn}{}_p
-g^2(\phi_mX^i_n\phi_q+2bX^j_mX^i_nX^j_q)f^{mn}{}_of^{oq}{}_p,
\nonumber\\
0&=&D^\a D_\a\phi_p-\frac{i}{2}g[(\bar\psi_{m+}\Gamma_5\psi_{n+})
-2b(\bar\psi_{m-}\Gamma_5\psi_{n-})] f^{mn}{}_p-2bg^2X^i_m\phi_nX^i_qf^{mn}{}_of^{oq}{}_p,
\nonumber\\
0&=&\Gamma^\a D_\a\psi_{m-}+g\Gamma^5\psi_{n-}\phi_pf^{np}{}_m+g\Gamma^5 \Gamma^i \psi_{n+}X^i_pf^{np}{}_m,\nonumber\\
0&=&\Gamma^\a D_\a\psi_{m+}-g\Gamma^5 \psi_{n+}\phi_pf^{np}{}_m+2bg\Gamma^5\Gamma^i\psi_{n-}X^i_pf^{np}{}_m,
\nonumber\\
0&=&gD_{[\a}H_{\b\g]5p}=D_{[\a}F_{\b\g]p}
\\
0&=&D^\a F_{\a\b p}-\frac{i}{2}g^2[(\bar\psi_{m+}\Gamma_\b
\psi_{n+})+2b(\bar\psi_{m-}\Gamma_\b
\psi_{n-})] f^{mn}{}_p-g^2(\phi_mD_\b\phi_n+2bX^i_mD_\b X^i_n)f^{mn}{}_p,
\nonumber\\
0&=&\partial_5\phi_{m}=\partial_5 X^i_{m}
=\partial_5\psi_{m-}=\partial_5\psi_{m+}=\partial_5 H_{\mu\nu\rho m}=\partial_5 g.\nonumber
\ee
Here the covariant derivative is defined as
\e
D_\a\phi_p=\partial_\a\phi_p+(A_\a)_m\phi_nf^{mn}{}_p.
\ee
It can be seen that the original equations of motion of H-fields are converted into two sets of equations: (1) the equations of motion of Yang-Mills fields in 5D; (2) the Bianchi identity for the field strength $F_{\b\g}$ in 5D.

And the supersymmetry transformations (\ref{susy4}) become
\e\label{susy7}
\d\phi_m&=&-i\bar\ep_+\psi_{m+},\nonumber\\
\d X^i_m&=&i\bar\ep_+\Gamma^i\psi_{m-},\nonumber\\
\d\psi_{m-}&=&\Gamma^\a\Gamma^i\ep_+D_\a X^i_m+g\Gamma_5\Gamma^i\ep_+  X^i_n\phi_p f^{np}{}_{m},\nonumber\\
\d\psi_{m+}&=&\Gamma^\a\ep_+D_\a\phi_m+\frac{1}{2g}
\Gamma_{\a\b}\Gamma_5\ep_+F^{\a\b}_m+bg\Gamma_5\Gamma^{ij}_{-}\ep_+ X^i_nX^j_p f^{np}{}_{m},
\nonumber\\
\d A_\a^m&=&ig\bar\ep_+\Gamma_{\a}\Gamma_5\psi^m_{+}.
\nonumber\\
\ee
These are the $\CN=1$ supersymmetry transformations associated with the 5D SYM theory. As observed by LP \cite{LP}, the coupling constant of the SYM theory is related to the constant $g$ as follows
\e
g=g^2_{YM}.
\ee
Notice that the continuous parameter $b$ still survives in the $\CN=1$ SYM theory.

The $\CN=1$ SYM theory can be the dual gauge theory of multi D4-branes. It would be interesting to study their large N limit and to construct their gravity duals.

Recall that in the special case of $b=\frac{1}{2}$, the supersymmetry of the 6D $(1,0)$ theory is promoted to $(2,0)$, and theory becomes the $(2,0)$ theory (see Section \ref{20theory}). Substituting the vev (\ref{vev}) into the equations (\ref{equations3}) and the supersymmetry transformations (\ref{susy6}) of the $(2,0)$ theory, one can obtain the maximally supersymmetric ($\CN=2$) Yang-Mills theory in 5D. For details, see Ref. \cite{LP}.

We end this section by commenting on the other possibilities:
$\langle C^\mu\rangle$ is a light-like vector or a time-like vector. In Ref. \cite{LP},
it was argued that if one uses the null reduction
\e
\langle C^\mu\rangle=g(1,0,\ldots,0,1),\quad\langle C^\mu \rangle\langle C_\mu \rangle=0,
\ee
i.e. $\langle C^\mu\rangle$ is a light-like vector,
 the $(2,0)$ theory can be used
to describe a system consisting of both M2 and M5-branes. Since the $(2,0)$ theory
is a special case of our $(1,0)$ theory, we expect that the $(1,0)$ theory can be
also a gauge description of some system containing M2 and M5-branes by choosing the
null reduction.
On the other hand, like its $(2,0)$ counterpart \cite{LR}, this $(1,0)$ theory may be also a light-cone description of multiple M5-branes.
However, we leave the work of exploring this particular 6D $(1,0)$ theory to the future.

Finally, if we choose $\langle C^\mu\rangle$ as the time-like vector
\e\label{time}
\langle C^\mu\rangle=g(1,0,\ldots,0),
\ee
all fields are static:
\e
0&=&D_0\phi_{m}=D_0 X^i_{m}
=D_0\psi_{m-}=D_0\psi_{m+}=D_0 H_{\mu\nu\rho m}=\partial_0 g,
\ee
and the theory may be used to describe static 5-branes in 11D \cite{LP}.
It would be nice to investigate this theory further.


\section{Acknowledgement}
We are grateful to Bin Chen for useful discussions. This work is supported by the Ren-Cai Foundation of Beijing Jiaotong University through Grant No. 2013RC029, and supported by the Scientific Research Foundation for Returned Scholars, Ministry of Education of China.

\appendix

\section{Conventions and Useful Identities}\label{Identities}
Following the idea of Ref. \cite{LP}, we will also work with 32-component Majorana fermions. However, in our case, these are $SO(9,1)$ Majorana fermions. The gamma matrices are real. Under the decomposition $SO(9,1)\rightarrow SO(5,1)\times SO(4)$, one can define the chirality matrix $\Gamma_{012345}$ of $SO(5,1)$ and the chirality matrix $\Gamma_{6789}$ of $ SO(4)$. The fermionic fields $\psi$ are anti-chiral with respect to $\Gamma_{012345}$, i.e.
\e\label{chiral1}
\Gamma_{012345}\psi=-\psi,
\ee
while the parameters of supersymmetry transformations $\ep$ are chiral,
\e\label{chiral2}
\Gamma_{012345}\ep=\ep.
\ee
We also define
\e\label{chiral4}
\psi_{\pm}=\frac{1}{2}(1\pm \Gamma_{6789})\psi\quad {\rm and}\quad\ep_{\pm}=\frac{1}{2}(1\pm \Gamma_{6789})\ep.
\ee
Under these definitions,  we have
\e\label{chiral3}
\Gamma_{6789}\psi_{\pm}=\pm\psi_{\pm} \quad {\rm and}\quad \Gamma_{6789}\ep_{\pm}=\pm\ep_{\pm}.
\ee

The product of two spinors is defined as
\e
\bar\psi\eta=\psi^TC\eta=\psi^T\Gamma_0\eta,
\ee
where $\eta$ and $\psi$ have opposite chiralities with respect to $\Gamma_{012345}$. We have chosen $\Gamma_0=-\Gamma^T_0$ as the charge conjugation matrix $C$. It obeys the equations
\e
C\Gamma^\mu C^{-1}=-\Gamma^{\mu T}\quad {\rm and}\quad C\Gamma^i C^{-1}=-\Gamma^{i},
\ee
where $\mu=0,\ldots,5$ and $i=6,\ldots,9$.

To work out the Fierz identities, it is convenient to introduce the eleventh gamma matrix
\e\label{gamma10}
\Gamma_{10}=\Gamma_{0}\Gamma_1\ldots\Gamma_9=\Gamma_{012345}\Gamma_{6789}.
\ee
Now the gamma matrices satisfy the Clifford algebra in eleven spacetime dimensions
\e
\{\Gamma_m, \Gamma_n\}=2\eta_{mn},
\ee
where $m=0, 1,\ldots, 10$, and $\eta_{mn}=$diag$(-, +,\ldots,+)$. The Fierz identity in 11D  reads \cite{LP}
\e\label{Fierz11d}
\ep_1\bar\ep_2=-2^{-5}\sum^5_{p=0}\frac{1}{p!}(-1)^{\frac{1}{2}(p-1)p}(\bar\ep_2\Gamma_{m_1\ldots m_p}\ep_1)\Gamma^{m_1\ldots m_p}.
\ee
The antisymmetric part of the above equation is
\e\label{Fierz11da}
\ep_1\bar\ep_2-\ep_2\bar\ep_1=-\frac{1}{16}\sum_{p=1,2,5}\frac{1}{p!}(-1)^{\frac{1}{2}(p-1)p}(\bar\ep_2\Gamma_{m_1\ldots m_p}\ep_1)\Gamma^{m_1\ldots m_p}.
\ee

Using Eqs. (\ref{chiral1}), (\ref{chiral2}), (\ref{chiral3}), and (\ref{gamma10}), one can reduce equation (\ref{Fierz11da}) to
\begin{equation}\label{Fierz2}
(\bar\ep_{2+}\psi_+)\ep_{1+}-(\bar\ep_{1+}\psi_+)\ep_{2+}=-\frac{1}{4}
(\bar\ep_{2+}\Gamma_\mu\ep_{1+})\Gamma^\mu\psi_+
-\frac{1}{192}(\bar\ep_{2+}\Gamma_{\mu\nu\lambda}
\Gamma^{ij}_-\ep_{1+})\Gamma^{\mu\nu\lambda}\Gamma^{ij}_-\psi_+,
\end{equation}
where $\mu, \nu, \lambda=0,\ldots, 5$ and $i, j=6,\ldots, 9$, and $\Gamma^{ij}_-$ are defined by the equations
\e\label{dual}
\Gamma^{ij}_\pm&=&\frac{1}{2}(\Gamma^{ij}\pm\frac{1}{2}\vp^{ijkl}\Gamma^{kl})\quad (\vp^{6789}=1),
\ee
satisfying $\Gamma^{ij}_\pm=\pm\frac{1}{2}\vp^{ijkl}\Gamma^{kl}_\pm$.
We see that $\Gamma^{kl}_{\pm}$ are the two sets of $SU(2)$ matrices\footnote{The $\pm$ signs carried by $\Gamma^{ij}_\pm$ are \emph{different} from those carried by $\ep_{\pm}$ (see Eqs. (\ref{chiral4}) and (\ref{dual})). We hope this will not cause any confusion.} of $SO(4)\cong SU(2)_L\times SU(2)_R$.
In deriving equation (\ref{Fierz2}), we have used the equation $\Gamma^{ij}_+\chi_+=0$.
This equation can be proved as follows: Multiplying both sides of the identity $\Gamma^{ij}=-\frac{1}{2}\vp^{ijkl}\Gamma^{kl}\Gamma_{6789}$ by $\chi_+$, and using  $\Gamma_{6789}\chi_+=\chi_+$, one obtains $\Gamma^{ij}\chi_+=-\frac{1}{2}\vp^{ijkl}\Gamma^{kl}\chi_+$. Using (\ref{dual}), one can see immediately that $\Gamma^{ij}_+\chi_+=0$. Similarly, one can prove that
$\Gamma^{ij}_-\chi_-=0$.

Similarly, using (\ref{Fierz11d}), we are able to derive the identity
\e\label{Fierz3}
\psi_{1-}\bar\psi_{2+}\nonumber
&=&\frac{1}{32}[(\bar\psi_{2+}\Gamma_\mu\Gamma^i\psi_{1-})
\Gamma^\mu\Gamma^i(1+\Gamma_{10})(1+\Gamma_{012345})\nonumber\\&&\quad\quad-\frac{1}{6}
(\bar\psi_{2+}\Gamma_{\mu\nu\rho}\Gamma^i\psi_{1-})
\Gamma^{\mu\nu\rho}\Gamma^i(1+\Gamma_{10})]
\ee
and the identity
\e\label{Fierz4}
\psi_{1+}\bar\psi_{2-}\nonumber
&=&\frac{1}{32}[(\bar\psi_{2-}\Gamma_\mu\Gamma^i\psi_{1+})
\Gamma^\mu\Gamma^i(1-\Gamma_{10})(1+\Gamma_{012345})\nonumber\\&&\quad\quad-\frac{1}{6}
(\bar\psi_{2-}\Gamma_{\mu\nu\rho}\Gamma^i\psi_{1+})
\Gamma^{\mu\nu\rho}\Gamma^i(1-\Gamma_{10})].
\ee

Also, with the definition of (\ref{gamma10}), we find that
\e\label{eppsi}
\Gamma_{10}\ep_{\pm}=\pm\ep_{\pm},\quad \Gamma_{10}\psi_{\pm}=\mp\psi_{\pm}.
\ee

Finally, we find that the following identity
\e\label{g1}
\Gamma_{\mu_1\ldots\mu_p}=\frac{(-1)^{\frac{1}{2}(p-1)p}}{(6-p)!}\vp_{\mu_1\ldots\mu_6}
\Gamma^{\mu_{p+1}\ldots\mu_6}\Gamma_{012345}
\ee
useful, where $\vp^{012345}=-\vp_{012345}=1$.

\end{document}